\documentclass[%
 preprint,
 superscriptaddress,
 amsmath,amssymb,
 aps,
 showkeys,
 titlepage,
 endfloats*.   %move all floats to the end of the document and put each on a separate page
]{revtex4-2}

\usepackage[utf8]{inputenc}
\usepackage[english]{babel}

\usepackage{graphicx}% Include figure files
\usepackage{dcolumn}% Align table columns on decimal point
\usepackage{bm}% bold math

\usepackage[colorlinks = true,
            linkcolor = blue,
            urlcolor  = blue,
            citecolor = red,
            anchorcolor = blue]{hyperref}

\begin{document}

%\preprint{APS/123-QED}

\title{Room-temperature Magnetic Thermal Switching by Suppressing Phonon-Magnon Scattering} 

\author{Fanghao Zhang}
\affiliation{%
Department of Mechanical Engineering, University of California, Santa Barbara, CA 93106-5070, USA}%

%feel free to add other authors

\author{Lokanath Patra}
\affiliation{%
Department of Mechanical Engineering, University of California, Santa Barbara, CA 93106-5070, USA}%

\author{Yubi Chen}
\affiliation{%
 Department of Physics, University of California, Santa Barbara, California 93106-9530, USA}%
\affiliation{%
Department of Mechanical Engineering, University of California, Santa Barbara, CA 93106-5070, USA}%

\author{Wenkai Ouyang}
\affiliation{%
Department of Mechanical Engineering, University of California, Santa Barbara, CA 93106-5070, USA}%

\author{Paul Sarte}
\affiliation{Material Research Laboratory, University of California, Santa Barbara, CA 93106, USA}

\author{Shantal Adajian}
\affiliation{%
Department of Mechanical Engineering, University of California, Santa Barbara, CA 93106-5070, USA}%

\author{Xiangying Zuo}
\affiliation{%
Department of Mechanical Engineering, University of California, Santa Barbara, CA 93106-5070, USA}%

\author{Runqing Yang}
\affiliation{%
Department of Mechanical Engineering, University of California, Santa Barbara, CA 93106-5070, USA}%

\author{Tengfei Luo}
\affiliation{%
Department of Aerospace and Mechanical Engineering, University of Notre Dame, Notre Dame, IN 46556, USA}%

\author{Bolin Liao}
\email{bliao@ucsb.edu} 
\affiliation{%
Department of Mechanical Engineering, University of California, Santa Barbara, CA 93106-5070, USA}%

%\date{\today}

\begin{abstract}
Thermal switching materials, whose thermal conductivity can be controlled externally, show great potential in contemporary thermal management.
Manipulating thermal transport properties through magnetic fields has been accomplished in materials that exhibit a high magnetoresistance. 
However, it is generally understood that the lattice thermal conductivity attributed to phonons is not significantly impacted by the magnetic fields.
In this study, we experimentally demonstrate the significant impact of phonon-magnon scattering on the thermal conductivity of the rare-earth metal gadolinium near room temperature, which can be controlled by a magnetic field to realize thermal switching.
Using first-principles lattice dynamics and spin-lattice dynamics simulations, we attribute the observed change in phononic thermal conductivity to field-suppressed phonon-magnon scattering. 
This research suggests that phonon-magnon scattering in ferromagnetic materials is crucial for determining their thermal conductivity, opening the door to innovative magnetic-field-controlled thermal switching materials.
\end{abstract}

%Use showkeys class option if keyword display desired
                            
\maketitle

%\tableofcontents

\section{Introduction}
Solid-state thermal switches with a tunable thermal conductivity have gained substantial attention in recent years due to their potential for addressing pressing challenges of thermal management in modern technologies \cite{wehmeyer2017thermal}.
Differing from traditional static thermal components, which typically are either thermally conductive or insulating, thermal switches offer a dynamic response to external stimuli, enabling transitions between thermally conductive and insulating states.
Realizing this inherent capacity of thermal switching not only contributes to our fundamental understanding of heat carrier interactions in condensed matter systems but also opens up technological avenues towards externally controllable heat flow. Such control is invaluable in a range of applications, from adiabatic demagnetization refrigeration (ADR) \cite{shirron2004portable,jahromi2014piezoelectric} to thermal management of electronic and energy systems \cite{hao2018efficient} and thermoelectric power generation \cite{atouei2018protection}.
Several approaches have been explored to achieve solid-state thermal switching with various mechanisms, leveraging different types of external stimuli, including manipulating phonon transport properties \cite{aryana2022observation,ihlefeld2015room,liu2023low,wooten2023electric} or interfacial chemical bonding and charge distribution~\cite{li2023electrically} using an external electric field, utilizing large magnetoresistance of electrons under a magnetic field~\cite{kimling2015spin}, phase transitions induced electrochemically~\cite{lu2020bi}, thermally~\cite{zhang2013high, shrestha2019high} or by light~\cite{shin2019light}, electrochemical intercalation~\cite{sood2018electrochemical,zhu2016tuning}, and manipulation of the displacement amplitude of atomic vibrations via hydration~\cite{tomko2018tunable}.
The underlying physics of these switching mechanisms can vary significantly, and the discovery of new switching mechanisms can both inspire novel forms of stimuli and guide the search for new thermal switching materials.

% The scientific exploration of thermal transport in solid-state materials has typically relied on understanding and tailoring the transport properties of heat carriers such as electrons and phonons. The ability to actively control these properties can generate observable thermal switching effects.
While previous thermal switch designs have relied mainly on electric fields and electrochemical processes, magnetic-field-driven thermal switching mechanisms remain less explored. The noncontact nature of magnetic fields can be advantageous for certain applications. Moreover, magnetic thermal switches can be particularly desirable in applications where large magnetic fields are already present, such as in ADR devices. In magnetic materials, heat conduction is determined by the transport and interaction processes of electrons, phonons, and magnons. Existing thermal switching studies have focused on the impact of a magnetic field on electron transport, namely magnetoresistance, essentially arising from the alteration of electron trajectories in the presence of the external magnetic field. Since electrons also carry heat,
materials exhibiting high magnetoresistance can achieve a substantial thermal switching ratio when an external magnetic field is applied \cite{hirata2023magneto}. 
In comparison, the impact of an external magnetic field on the phononic transport properties is not as direct. Jin \textit{et al.} showed that a diamagnetic force induced by lattice vibrations can lead to small magnetic-field-dependent changes in the phononic thermal conductivity of nonmagnetic materials at cryogenic temperatures~\cite{jin2015phonon}.
Recent computational studies have suggested that an external field can exert an indirect influence on phonons through phonon-magnon coupling~\cite{stockem2018anomalous}.
In magnetic materials, a collective excitation in the spin structure, known as a magnon, can contribute to heat conduction at cryogenic temperature in magnetic materials \cite{kittel2018introduction,douglass1963heat,jin2015phonon,boona2014magnon}.
Although, at room temperature, magnons contribute minimally to the overall thermal conductivity due to their short mean free path and small heat capacity~\cite{jacobsson1989thermal}, they can effectively scatter electrons and phonons and thereby indirectly modulate thermal conductivity.
Previous research has investigated the phonon-magnon interaction at cryogenic temperatures, where their impact on thermal conductivity is appreciable \cite{bhandari1966scattering,sanders1977effect}. 
However, the possibility of realizing solid-state thermal switching via controlling phonon-magnon interaction with an external magnetic field has not been explored, especially near room temperature.
It is anticipated that materials demonstrating strong spin-lattice interactions, which could result in sufficiently frequent phonon-magnon scatterings comparable to phonon-phonon scatterings, can display significant thermal switching when exposed to changes in external magnetic fields~\cite{sharma2004thermal}.

For this purpose, a suitable candidate material is elemental gadolinium (Gd), a rare-earth metal known to possess strong spin-lattice coupling \cite{patra2023indirect}.
The thermal conductivity of Gd has been a subject of investigation since 1964, when Arajs \textit{et al.} measured thermal and electric conductivity of Gd without an external magnetic field. Their study highlighted the pivotal role of the magnetic transition in Gd's thermal transport \cite{arajs1964thermal}. Further work by Jacobsson \textit{et al.}  concluded that magnons contribute to heat conduction in Gd below its Curie temperature ($T_c$=293\,K) and that the lattice thermal conductivity strongly depends on pressure \cite{jacobsson1989thermal}.
In a study conducted by Glorieux \textit{et al.}, the thermal conductivity of Gd was measured under a modest external magnetic field (60\,mT) using a photoacoustic method and was shown to change with the field strength, especially near $T_c$. The authors suggested that the observed changes in thermal conductivity with the applied field could be attributed to two main factors: an increase in the effective mass of conduction electrons caused by localized spins, and alterations in the density of states at the Fermi surface due to the indirect Ruderman–Kittel–Kasuya–Yosida (RKKY) exchange interaction, which is particularly significant in rare-earth materials \cite{glorieux1995photoacoustic, jensen1991rare}.
However, an important aspect that was overlooked in their study was the influence of the magnetic field on phononic thermal conductivity by modifying spin-lattice coupling. 

In this work, we employed a customized steady-state measurement system to measure thermal and electrical conductivity of a single-crystal Gd sample at varying temperatures under an external magnetic field up to 9 T, in order to clarify the true origin of its magnetic-field-dependent thermal conductivity.
We found that the thermal conductivity of Gd increases with the magnetic field strength, specifically in the vicinity of $T_c$, while the change decreases at temperatures away from $T_c$.
By isolating the electronic and phononic contributions to the thermal conductivity, we found that the phononic contribution plays a crucial role in the observed increase in the total thermal conductivity. 
This conclusion was further corroborated by first-principles spin-lattice dynamic (SLD) simulation, which showed that phonon-magnon scatterings play an important role in thermal transport in Gd even at room temperature. The external magnetic field reduces the magnon population and suppresses phonon-magnon scattering, leading to the increase in the thermal conductivity.
Our findings provide important fundamental insights into magnetic-field-dependent thermal transport in magnetic materials and suggest that controlling phonon-magnon scattering through an external magnetic field in materials with a strong spin-lattice coupling can be a promising new mechanism for solid-state thermal switching. Because materials with strong spin-lattice coupling often show a prominent magnetocaloric effect~\cite{patra2023indirect}, our study suggests that further exploration of known magnetocaloric materials may lead to promising candidates for thermal switching applications.

\section{Results and Discussions}
\begin{figure*}[!htb]
\includegraphics[width=1\textwidth]{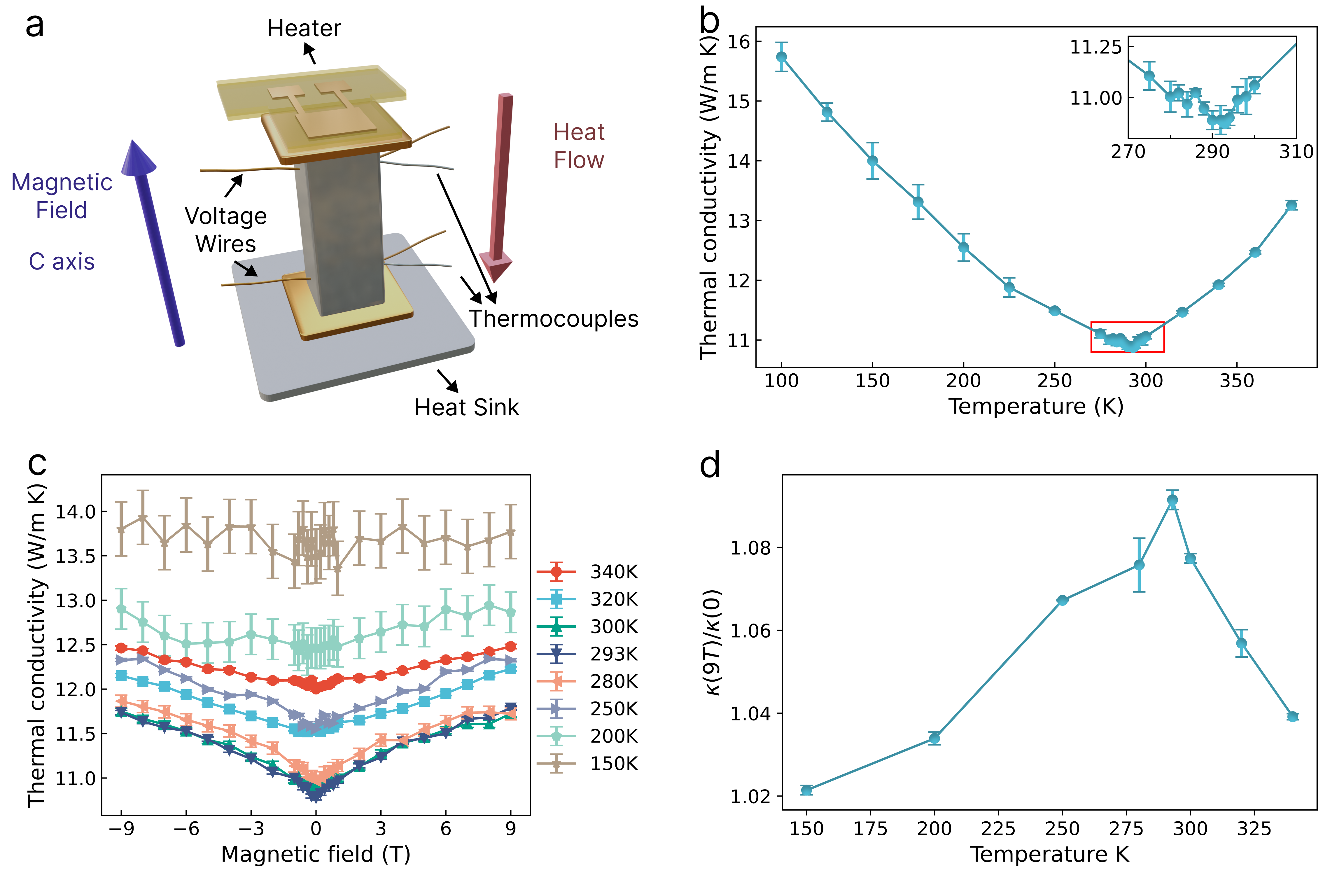}
\caption{(a) The schematic of the experimental setup. (b) Measured thermal conductivity of single crystalline Gd as a function of temperature without an external field. The inset shows the data near the Curie temperature. The thermal conductivity shows a minima around its Curie temperature (293 K). (c) Measured thermal conductivity of single crystalline Gd as a function of the external magnetic field at different temperatures. The external magnetic field increases the thermal conductivity near the Curie temperature. (d) The switching ratio of the thermal conductivity of Gd under a 9-T magnetic field. The error bars include uncertainties due to systematic and random errors as discussed in the Supplementary Information.} 
\label{fig:fig1}
\end{figure*}

The thermal conductivity of single crystalline Gd was measured using a customized steady-state method setup, as shown in Fig.~\ref{fig:fig1}a, with further details provided in the Supplementary Information and Fig.~S1. The magnetic field is applied along the hexagonal c-axis. The thermal conductivity and the electrical resistivity are also measured along the c-axis.
As shown in Figure~\ref{fig:fig1}b, the temperature-dependent thermal conductivity in the absence of an applied magnetic field exhibits a trend consistent with previously reported data \cite{jacobsson1989thermal}. 
Specifically, the thermal conductivity decreases from 15.8 W/m·K at 100 K to approximately 10.8 W/m·K around its Curie temperature $T_c$ of 293 K, a reduction attributable to increased phonon-phonon scattering. 
Notably, the observed increase in thermal conductivity above $T_c$ deviates from the typical behavior expected in crystalline solids, suggesting particularly strong scatterings of heat carriers in Gd near the magnetic phase transition at $T_c$.

Magnetic-field-dependent thermal conductivity measurements were conducted at various temperatures, both near (280, 293, and 300 K) and significantly distant (150, 200, 250, 320, and 340 K) from the Curie temperature, with magnetic field strengths of up to 9 T, as illustrated in Fig.~\ref{fig:fig1}c. 
Near $T_c$, the thermal conductivity increases with the strength of the external magnetic field, rising from 10.75 W/m·K with zero field to 11.6 W/m·K under a 9-T magnetic field. 
This field dependence decreases at temperatures away from $T_c$ and diminishes at 150 K. Consequently, as illustrated in Fig.~\ref{fig:fig1}d, the thermal conductivity switching ratio attains a maximum of 1.09 at the Curie temperature.

Thermal transport in metallic Gd can be mediated by electrons, phonons, and magnons. Since the population of magnons is suppressed by the magnetic field \cite{liao2014generalized}, the magnonic thermal conductivity is expected to decrease with the field, contrary to our observation. Furthermore, previous studies have suggested that the contribution of magnons in Gd at room temperature is negligible~\cite{jacobsson1989thermal}. Therefore, we focus on analyzing the field dependence of electronic and phononic thermal conductivity. 

% significantly influenced by conduction electrons, a phenomenon extensively documented in previous research \cite{jensen1991rare}.
% The Wiedemann-Franz law offers a robust framework for understanding the correlation between electronic thermal conductivity and electrical conductivity in such materials. \cite{kittel2018introduction}
% However, discrepancies observed between the measured thermal conductivity and the calculated electronic thermal conductivity in Gadolinium suggest the involvement of additional heat carriers, such as phonon and magnon. \cite{nellis1969thermal}
% While it is acknowledged that magnon contributions to thermal conductivity can be substantially subdued under strong external magnetic fields at low temperatures, this suppression is not as effective at room temperature.  
% Despite this, magnon, particularly those with relatively low energies, maintain a a palpable occupancy at room temperature. \cite{pan2023ab}

\begin{figure*}[!htb]
\includegraphics[width=1\textwidth]{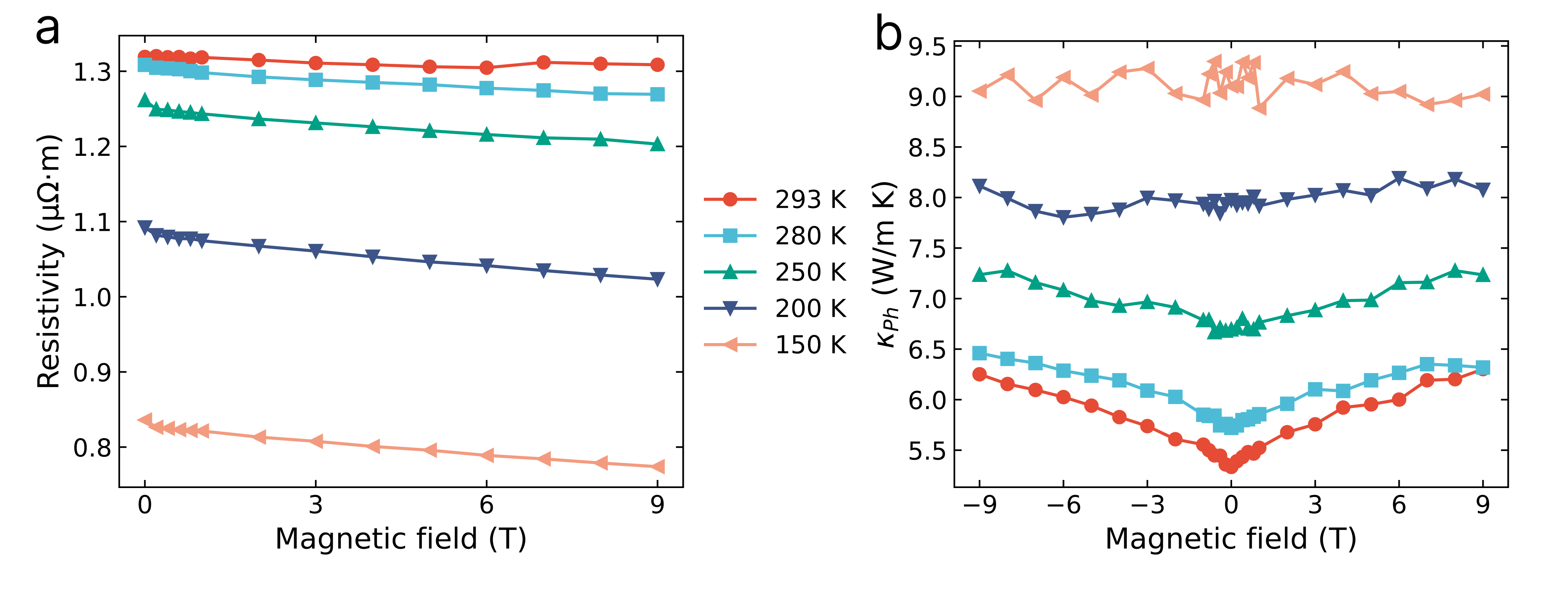}
\caption{(a) The measured electrical resistance of single crystalline Gd as a function of an external magnetic field at different temperatures. A negative magnetoresistance is measured at all temperatures. (b) The phononic contribution to the thermal conductivity of single crystalline Gd as a function of an external magnetic field at different temperatures. The phononic contribution is estimated by subtracting the electronic contribution from the total thermal conductivity using the Wiedemann-Franz law. } 
\label{fig:fig2}
\end{figure*}

To estimate the electronic contribution to the thermal conductivity, the magnetoresistance of the same Gd sample was measured. The electrical resistance of the sample as a function of temperature without an external field is provided in Fig.~S2a, showing a typical metallic behavior.
As depicted in Fig.~\ref{fig:fig2}a, the sample exhibits a linear negative magnetoresistance, a result of suppressed spin-electron scattering as the local magnetic moments are aligned by the applied external magnetic field, in agreement with previous reports \cite{hiraoka1971magnetoresistance,mcewen1977magnetoresistance}. In particular, we found that the magnetoresistance is higher at lower temperature compared to that near the Curie temperature, a similar trend reported previously along the c-axis~\cite{hiraoka1971magnetoresistance}. This behavior is in contrast to the observed magnetic-field dependence of the thermal conductivity, which is maximized near the Curie temperature.
 % This phenomenon can be explained by the increased mobility of conduction electrons in the ferromagnetic state, facilitated by their spin alignment with the uniform magnetic field, which reduces electrical resistance. However, as the temperature increases towards $T_c$, the interaction between conduction electrons and local spins intensifies, weakening the effectiveness of the external magnetic field in suppressing this scattering.
Based on the Wiedemann-Franz law, which remains applicable in a magnetic field \cite{ziman1960electrons}, the electronic thermal conductivity can be calculated from the magnetoresistance results. Using the electronic band structure of Gd calculated from first-principles with the density functional theory (DFT), the Lorenz number in Gd is calculated to be very close to the classical value of $2.44 \times 10^{-8}\,\mathrm{V}^2\,\mathrm{K}^{-2}$ in the temperature range of our experiment (Fig.~S3).
The calculated electronic thermal conductivity based on this Lorenz number is shown in Fig.~S2b. At $T_c$, the electronic thermal conductivity changes from 5.14 W/m·K with zero field to 5.26 W/m·K with a 9-T field. The phononic thermal conductivity can then be isolated by subtracting the electronic contribution from the total thermal conductivity. The result is
depicted in Fig.~\ref{fig:fig2}b. Similar to the total thermal conductivity, the magnetic-field-induced change in the phononic thermal conductivity peaks near the Curie temperature $T_c$. This result is in sharp contrast to the work by Glorieux \textit{et al.}\cite{glorieux1995photoacoustic}, where the field dependence was fully attributed to the electronic contribution.

Typically, phonons are not expected to directly interact with external magnetic fields due to their non-magnetic nature. 
However, it has been proposed that phonons can indirectly respond to such fields through spin-lattice coupling, affecting either phonon-phonon \cite{jin2015phonon} or phonon-magnon scatterings \cite{vu2023magnon,gofryk2014anisotropic}.
Given prior simulations that show only slight variations in the phonon dispersion relation in Gd under magnetic fields \cite{patra2023indirect}, we hypothesize that the marked field dependence of the phononic thermal conductivity is mainly attributable to the reduction in phonon-magnon scattering instead of changes in the phonon group velocity, phononic heat capacity, or phonon-phonon scattering. Below the Curie temperature $T_c$, the thermal fluctuations of the local magnetic moments are suppressed by the external magnetic field because the moments tend to align with the field. Equivalently, the magnon population is reduced by the external magnetic field. This can be seen from the equilibrium Bose-Einstein distribution that the magnons obey~\cite{liao2014generalized}:
\begin{equation}
    f_0(\omega)=\frac{1}{\exp(\frac{\hbar\omega+g\mu_B B}{k_B T})-1},
\end{equation}
where $\omega$ is the magnon frequency, $g$ is the Land\'{e} $g$ factor, $\mu_B$ is the Bohr magneton, $B$ is the external magnetic field, $k_B$ is the Boltzmann constant, and $T$ is the temperature. This effect is also reflected in the quenching of the heat capacity by the applied magnetic field (Fig.~S4). Thus, the reduction of the magnon population due to an external magnetic field leads to reduced scattering of heat-carrying phonons by magnons, giving rise to an increased phononic thermal conductivity with an applied field. In particular, 
this effect is maximized near the Curie temperature $T_c$, where the thermal fluctuations and the exchange interactions of the spins approximately balance each other. As a result, the application of a magnetic field near $T_c$ produces the largest change in the magnetization and, thus, the magnon population~\cite{dan1998magnetic}. The strong spin fluctuation and the resulting significant phonon-magnon scattering can also be responsible for the minimum in the thermal conductivity of Gd near $T_c$, as shown in Fig.~\ref{fig:fig1}b. 

% are not fully aligned, but applying an external magnetic field tends to enhance their alignment. 
% This magnetic field can induce phase transitions in Gadolinium, elevating $T_c$ to room temperature \cite{glorieux1995photoacoustic}, as reflected in the heat capacity data presented in Figure S2. 
% Beyond $T_c$, the heat capacity reaches a plateau under fields exceeding 2 Tesla.
% At temperatures significantly lower than $T_c$, both magnon and phonon densities are low, thereby limiting their scattering interactions. 
% Consequently, the suppression of these interactions is less effective, leading to a diminished field dependence of phonon thermal conductivity, particularly at temperatures like 150 K and 200 K. 
% As the temperature increases, the phonon density rises, and phonon scattering becomes the dominant mechanism influencing phononic thermal conductivity. 
% Thus, mitigating phonon-magnon interactions can substantially enhance thermal conductivity near the transition temperature, where electron-magnon scattering is less significant.

\begin{figure*}[!htb]
\includegraphics[width=1\textwidth]{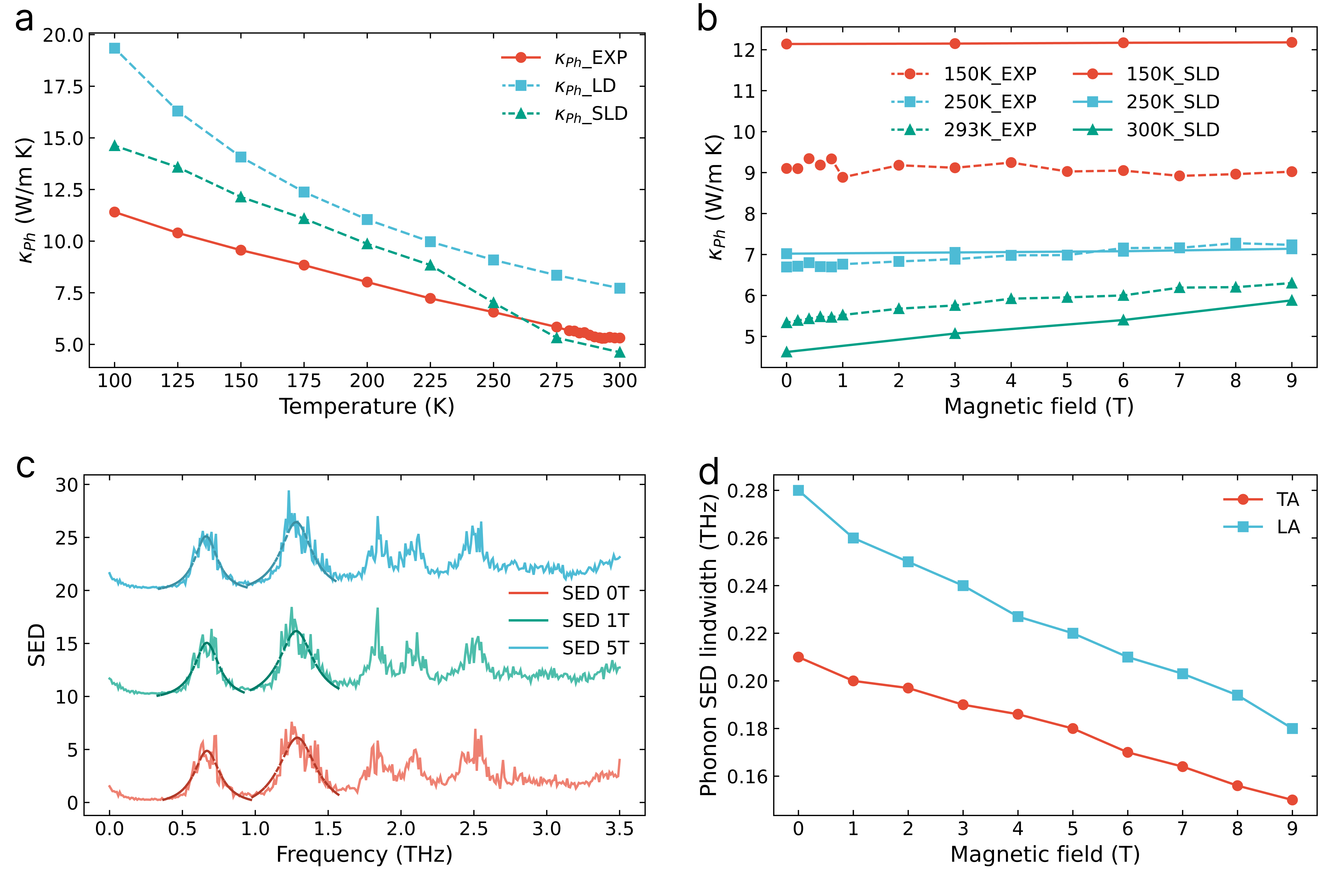}
\caption{(a) Comparison of the experimentally measured phononic thermal conductivity of singal crystalline Gd without an external magnetic field ($\kappa_{Ph}$\_EXP) to the results from first-principles lattice dynamics ($\kappa_{Ph}$\_LD) and spin-lattice dynamics ($\kappa_{Ph}$\_SLD) simulations. (b) Comparison of the experimentally measured magnetic field dependence of the phononic thermal conductivity in single crystalline Gd to the spin-lattice dynamics simulation results at different temperatures. (c) The phonon spectral energy density (SED) evaluated at $\mathbf{q}=[0.25,0,0]$ from the SLD simulation as a function of an external magnetic field. The two lowest peaks, corresponding to the TA and LA acoustic modes, are fitted to a Lorentzian function to extract their linewidths. (d) The extracted peak linewidth of the TA and LA modes in single crystalline Gd from the SLD simulation as a function of an external magnetic field up to 9 T. The observed narrowing of the SED peak linewidth in the presence of an external magnetic field suggests suppression of phonon scattering by the field.} 
\label{fig:fig3}
\end{figure*}

To verify our hypothesis and elucidate the mechanisms underlying the modulation of phononic thermal conductivity, we have employed first-principles lattice dynamics (LD) and spin-lattice dynamic (SLD) simulations to quantitatively explain our experimental results. Details of the calculations are summarized in the Supplementary Information. Briefly, in the LD approach~\cite{broido2007intrinsic}, the interatomic force constants (IFCs) are calculated at 0 K using DFT for the ferromagnetic ground state of Gd. These IFCs are used to compute the phonon-dispersion relation and phonon-phonon scattering rates at finite temperatures, from which the phononic thermal conductivity can be calculated. Within this approach, the spin degree of freedom is frozen and the phonon-magnon scattering is not captured. The calculated phononic thermal conductivity based on the LD simulation is shown in Fig.~\ref{fig:fig3}a. As expected, the LD simulation provides a theoretical maximum value that is significantly higher than our experimental result, indicating the potentially important role of phonon-magnon scattering. In contrast, in the SLD simulation, both the spin and the lattice degrees of freedom are allowed to fluctuate at a given temperature, with the spin exchange coefficients and interatomic potentials obtained from first-principles using DFT. With the SLD approach, the phonon-magnon scattering is fully captured by taking into account the dependence of the spin exchange interaction on the interatomic distance~\cite{ma2008large,patra2023indirect}.   
As shown in Fig.~\ref{fig:fig3}a, when incorporating phonon-magnon interactions through the SLD simulation, the calculated $\kappa_{ph}$ aligns more closely with the experimental data. The small discrepancy between SLD and the experimental results might be due to defect-scatterings in the sample. The difference between the LD and SLD results indicates the contribution of phonon-magnon scattering in the determination of the thermal conductivity of Gd.
Moreover, the SLD simulation also successfully and effectively replicates the experimental field dependence observed in the phononic thermal conductivity, as shown in Fig.~\ref{fig:fig3}b. In particular, the maximum switching effect near $T_c$ is captured in the SLD simulation. To obtain a microscopic understanding of the field-suppressed phonon-magnon scattering, we calculated
the peak linewidth of the phonon spectral energy density (SED) at $T_c$ under different external magnetic fields from the trajectories of the ionic motion in the SLD simulation~\cite{wu2018magnon}. The peak linewidth is inversely proportional to the lifetime of the corresponding phonon modes, including contributions from both phonon-phonon and phonon-magnon scatterings. As shown in Fig.~\ref{fig:fig3}c and d, the peak linewidths associated with the transverse acoustic (TA) and the longitudinal acoustic (LA) phonon modes decrease monotonically with the applied external magnetic field. This result provides direct evidence that the total phonon scattering rates are reduced as a result of the applied magnetic field, leading to increased phononic thermal conductivity.

\section{Conclusion}
In summary, we utilized a customized steady-state measurement system to investigate the thermal and electrical magneto-conductivity properties of single crystalline Gd at various temperatures. 
Our findings reveal a notable increase in the phononic thermal conductivity with an external magnetic field, a phenomenon particularly pronounced near the Curie temperature. 
To explore the origin of this field dependence, we implemented first-principles LD and SLD simulations. 
These simulations pointed to the suppression of phonon-magnon scattering as the primary factor contributing to the observed increase in the phononic thermal conductivity.
Our study paves the way for the development of innovative thermal switching materials driven by an external magnetic field, especially those exhibiting strong spin-lattice interactions, opening new possibilities in the field of material science and thermal engineering.

\begin{acknowledgments}
The authors acknowledge the help with the SED analysis from Xufei Wu. This work is based on research supported by the U.S. Office of Naval Research under the award number N00014-22-1-2262. The development of the spin-lattice dynamics simulation is supported by the National Aeronautics and Space Administration (NASA) under award number 80NSSC21K1812.
Y.C. also acknowledges the support from the Graduate Traineeship Program of the NSF Quantum Foundry via the Q-AMASE-i program under award number DMR-1906325 at the University of California, Santa Barbara (UCSB). 
This work used Stampede2 at Texas Advanced Computing Center (TACC) and Expanse at San Diego Supercomputer Center (SDSC) through allocation MAT200011 from the Advanced Cyberinfrastructure Coordination Ecosystem: Services \& Support (ACCESS) program, which is supported by National Science Foundation grants 2138259, 2138286, 2138307, 2137603, and 2138296. Use was also made of computational facilities purchased with funds from the National Science Foundation (CNS-1725797) and administered by the Center for Scientific Computing (CSC). The CSC is supported by the California NanoSystems Institute and the Materials Research Science and Engineering Center (MRSEC; NSF DMR 2308708) at UCSB.

\end{acknowledgments}

\end{document}